\documentclass[conference]{IEEEtran}
\IEEEoverridecommandlockouts
\usepackage{cite}
\usepackage{amsmath,amssymb,amsfonts}
\usepackage{algorithmic}
\usepackage{graphicx}
\usepackage{textcomp}
\usepackage{xcolor}
\def\BibTeX{{\rm B\kern-.05em{\sc i\kern-.025em b}\kern-.08em
    T\kern-.1667em\lower.7ex\hbox{E}\kern-.125emX}}
    
\usepackage{numprint}    
    
\usepackage{acronym}
\newacro{sme}[SME]{Small and Medium-sized Enterprise}
\newacro{it}[IT]{Information Technology}
\newacro{ot}[OT]{Operation Technology}
\newacro{cps}[CPS]{Cyber-Physical System}
\newacro{ids}[IDS]{Intrusion Detection System}
\newacro{cpps}[CPPS]{Cyber-Physical Production System}
\newacro{ids}[IDS]{Intrusion Detection System}
\newacro{svm}[\textit{SVM}]{\textit{Support Vector Machine}}
\newacro{wsn}[WSN]{Wireless Sensor Network}
\newacro{darpa}[DARPA]{Defense Advanced Research Projects Agency}
\newacro{kdd}[KDD]{Knowledge Discovery in Databases}
\newacro{scada}[SCADA]{Supervisory Control And Data Acquisition}
\newacro{dpi}[DPI]{Deep Packet Inspection}
\newacro{dmz}[DMZ]{De-Militarized Zone}
\newacro{manet}[MANET]{Mobile Ad hoc NETwork}
\newacro{svm}[\textit{SVM}]{\textit{Support Vector Machine}}
\newacro{iiot}[IIoT]{Industrial Internet of Things}
\newacro{mass}[\textit{MASS}]{\textit{Mueen's Algorithm for Similarity Search}}
\newacro{opcua}[\textit{OPC UA}]{\textit{Object Linking and Embedding for Process Control Unified Architecture}}
\newacro{hmi}[HMI]{Human Machine Interface}
\newacro{mtu}[MTU]{Master Terminal Unit}
\newacro{plc}[PLC]{Programmable Logic Controller}
\newacro{mcar}[MCAR]{Missing Completely At Random}
\newacro{pca}[\textit{PCA}]{\textit{Principal Component Analysis}}
 \newacro{fp}[\textit{FP}]{\textit{False Positive}}
\newacro{tp}[\textit{TP}]{\textit{True Positive}}
\newacro{fn}[\textit{FN}]{\textit{False Negative}}
\newacro{tn}[\textit{TN}]{\textit{True Negative}}

\usepackage{booktabs}

\usepackage[hidelinks,bookmarks=false]{hyperref}
    
\begin{document}

\title{Anomaly-based Intrusion Detection in Industrial Data with \textit{SVM} and \textit{Random Forests}
\thanks{This work has been supported by the Federal Ministry of Education and Research of the Federal Republic of Germany (Foerderkennzeichen 16KIS0932, IUNO Insec).
The authors alone are responsible for the content of the paper. This is a preprint of a work accepted but not yet published at the 27th International Conference on Software, Telecommunications and Computer Networks (SoftCOM). Please cite as: S.D. Duque Anton, S. Sinha, H.D. Schotten: ``Anomaly-based Intrusion Detection in Industrial Data with SVM and Random Forests''. In: \textit{27th International Conference on Software, Telecommunications and Computer Networks (SoftCOM)}, IEEE, 2019}
}

\author{\IEEEauthorblockN{1\textsuperscript{st} Simon D. Duque Anton}
\IEEEauthorblockA{\textit{Intelligent Networks Research Group} \\
\textit{German Research Center for AI}\\
Kaiserslautern, Germany\\
simon.duque\_anton@dfki.de}
\and
\IEEEauthorblockN{2\textsuperscript{nd} Sapna Sinha}
\IEEEauthorblockA{\textit{Intelligent Networks Research Group} \\
\textit{German Research Center for AI}\\
Kaiserslautern, Germany\\
sapna.sinha@dfki.de}
\and
\IEEEauthorblockN{3\textsuperscript{rd} Hans Dieter Schotten}
\IEEEauthorblockA{\textit{Intelligent Networks Research Group} \\
\textit{German Research Center for AI}\\
Kaiserslautern, Germany\\
hans\_dieter.schotten@dfki.de}
}

\maketitle

\begin{abstract}
Attacks on industrial enterprises are increasing in number as well as in effect.
Since the introduction of industrial control systems in the 1970's,
industrial networks have been the target of malicious actors.
More recently,
the political and warfare-aspects of attacks on industrial and critical infrastructure are becoming more relevant.
In contrast to classic home and office IT systems,
industrial IT,
so-called OT systems,
have an effect on the physical world.
Furthermore,
industrial devices have long operation times,
sometimes several decades.
Updates and fixes are tedious and often not possible.
The threats on industry with the legacy requirements of industrial environments creates the need for efficient intrusion detection that can be integrated into existing systems.
In this work,
the network data containing industrial operation is analysed with machine learning- and time series-based anomaly detection algorithms in order to discover the attacks introduced to the data.
Two different data sets are used,
one \textit{Modbus}-based gas pipeline control traffic and one \textit{OPC UA}-based batch processing traffic.
In order to detect attacks,
two machine learning-based algorithms are used,
namely \textit{SVM} and \textit{Random Forest}.
Both perform well,
with  \textit{Random Forest} slightly outperforming \textit{SVM}.
Furthermore,
extracting and selecting features as well as handling missing data is addressed in this work.
\end{abstract}


\section{Introduction}
\label{sec:intro}
Over the last two decades,
industrial systems have increasingly become the target of malicious actors~\cite{Duque_Anton.2017a}.
In the 1970's,
electronic components were introduced to automate control systems,
followed by \ac{scada} systems that allowed for reuse and reprogramming of components.
At that time, 
security has not been an issue due to two reasons~\cite{Igure.2006}:
First,
industrial networks, 
also known as \ac{ot},
are physically separated from \ac{it} networks.
Second,
\ac{ot} networks are highly application specific,
an attacker supposedly has no chance of understanding and influencing the processes.
However,
the rise of the \ac{iiot} and Industry 4.0 rendered both assumptions obsolete.
The fourth industrial revolution creates novel use cases,
e.g. the digital factory~\cite{STEF2013451},
mostly based on the increased capabilities in communication and embedded intelligence.
However,
these use cases require interconnectivity and distribution of data,
inherently breaking the above-mentioned assumptions.
This drastically increases the attack surface,
especially since most industrial communication protocols were not designed with security in mind.
Many of them, 
such as \textit{Modbus}~\cite{Modbus.2012, ModbusIDA.2006} \textit{Profinet}~\cite{PROFIBUS.2017},
do not contain means for authentication or encryption.
As a consequence,
any attacker that has successfully broken the perimeter and moved laterally into the \ac{ot} network can listen and participate in the communication.
Breaking the perimeter of \ac{it} networks employed in the office area of industrial enterprises becomes the biggest hurdle for attackers and success in doing so leads to severe consequences,
e.g. the blackout in the Ukrainian power grid in December of 2015,
caused by a malware called \textit{Industroyer} or \textit{Crashoverride}~\cite{Cherepanov.2017, Dragos.2016, Lee.2016}.
Due to characteristics of \ac{ot} networks,
such as legacy devices,
but also periodic and unique behaviour,
\acp{ids} from the \ac{it} world cannot be transferred as they are.
Adaptions have to be made,
e.g. in creating industrial firewalls.
In this work,
methods based on machine learning as well as time series analysis are evaluated with respect to their capabilities to detect attacks in \ac{ot} networks.
The remainder of this work is structured as follows.
An overview of related work is provided in Section~\ref{sec:sota}.
The data sets used for the evaluation are introduced in Section~\ref{sec:data},
the algorithms implemented to detect the attacks are presented in Section~\ref{sec:algos}.
In Section~\ref{sec:eval},
the results are presented and discussed.
Finally,
this work is concluded in Section~\ref{sec:concl}.

\section{Related Work}
\label{sec:sota}
Due to the relevance of industrial enterprises to the national wealth of industrial nations and the severe potential impacts of \acp{cpps} on the physical world,
industrial intrusion detection has gained relevance in the research community.
\textit{Duque Anton et al.} have evaluated industrial network data with classifiers~\cite{Duque_Anton.2018b} as well as with time series-based methods~\cite{Duque_Anton.2018c}.
A survey regarding existing risk assessment methods is performed by \textit{Cherdantseva et al.},
as well as an assessment of the applicability of these methods in \ac{scada} scenarios~\cite{Cherdantseva.2016}.
\textit{Zhu et al.} provide an extensive overview of industrial \acp{ids} along several dimensions and compare security in \ac{it} and \ac{ot} networks~\cite{Zhu.2010}.
\textit{Garcia-Teodoro et al.} summarise challenges and solutions to anomaly detection in industrial networks~\cite{Garcia-Teodoro.2009}.
Attacks on industrial networks are analysed and evaluated by \textit{Caselli et al.}~\cite{Caselli.2015}.
\textit{Khalili and Sami} employ an a priori algorithm to detect attacks in industrial networks,
based on their regularity and knowledge about critical states~\cite{Khalili.2015}.
They focus on the timing behaviour of communication sequences that is expected to be regular in industrial environments.
\textit{Gao and Morris} address the detection of attacks in \textit{Modbus} communication with the help of an attack classification and terminology~\cite{Gao.2014}.
Another approach for detection in the legacy \textit{Modbus} communication is introduced by \textit{Morris et al.}~\cite{Morris.2012}.\\ \par
Wireless communication technologies are an enabler of novel business and application use cases for industry,
although wireless systems contain inherent attack surfaces.
A survey of intrusion detection in \acp{wsn} is performed by \textit{Butun et al.}~\cite{Butun.2014}.
\acp{manet} are a certain kind of wireless network,
addressed by \textit{Shakshuki et al.}~\cite{Shakshuki.2013}.
\textit{Wei et al.} employ a prediction-based intrusion detection system for wireless industrial networks~\cite{Wei.2013}.
\textit{Shin et al.} present methods of intrusion detection for such networks~\cite{Shin.2010},
commonly known as \acp{wsn}.
\textit{Zhang et al.} apply methods of intrusion detection to mobile networks~\cite{Zhang.2003}.
A roadmap for the use of machine learning-based anomaly detection methods in industrial networks is provided by \textit{Meshram and Haas}~\cite{Meshram.2017}.
\textit{Schuster et al.} propose a method for anomaly detection mechanisms based on learning a notion of normal behaviour~\cite{Schuster.2013}.
An evaluation of several machine learning methods for detecting malicious activities in \ac{scada} communication is provided by \textit{Beaver et al.}~\cite{Beaver.2013}.
\textit{Mukkamala et al.} analyse the \ac{darpa} \ac{kdd} cup '99 with neural networks and \acp{svm} in order to detect the attacks contained in the data set~\cite{Mukkamala.2002}.

\section{Data Set}
\label{sec:data}
In this work,
two distinct data sets are analysed with respect to detecting attacks.
The data sets are derived from different sources as well as processes and employ different protocols,
namely \textit{Modbus} and \ac{opcua}.
In the remainder of this work,
the \textit{Modbus}-based data set is referred to as \textit{DS1},
the \ac{opcua}-based one as \textit{DS2}.
An overview of both data sets can be found in Table~\ref{tab:data_sets_overview}.
\begin{table}[h!]
\renewcommand{\arraystretch}{1.3}
\caption{Data Sets Analysed in this Work}
\label{tab:data_sets_overview}
\centering
\scriptsize
\begin{tabular}{l l r r r r}
\toprule
\textbf{ID} & \textbf{Protocol} & \textbf{No. of Packets} & \textbf{Duration} & \textbf{Attacks} & \textbf{No. of mal. Packets} \\
\textit{DS1} & \textit{Modbus} & \numprint{274627} & 4 d &  & \numprint{60048} \\
\textit{DS2} & \ac{opcua} & \numprint{4910} & 41 m & 2 & \numprint{702} \\
\bottomrule
\end{tabular}
\end{table}

\subsection{\textit{Modbus}-based Data Set \textit{DS1}}
The \textit{Modbus}-based data set has been presented by \textit{Morris et al.}~\cite{Morris.2015}.
A gas pipeline has been set up in a simulation environment.
This data set is an extension of a previous work of \textit{Beaver et al.}~\cite{Beaver.2013} which contained attacks that have been easily discovered by machine learning algorithms.
The goal of this data set was to increase complexity and realism.
Overall,
the simulations consists of four parts:
A \ac{hmi},
a \ac{plc},
a network simulation and a virtual process.
It is shown that an \ac{hmi} and a \ac{mtu} respectively are used to control a control logic that,
in turn,
influences a pump and valve in a physical environment.
In order to follow a real physical process as closely as possible,
the \textit{Matlab Simulink} package \textit{SimHydraulics} (now called \textit{Simscape Fluids})~\cite{mathworks.} was used to simulate a pump,
a valve,
a pipeline and a fluid as well as its flow.
The simulation has been run for the duration of four days,
creating \numprint{274627} packets,
containing 20 features each.
In Table~\ref{tab:modbus_features},
the complete feature list is shown.
\begin{table}[h!]
\renewcommand{\arraystretch}{1.3}
\caption{Features Available in \textit{DS1}}
\label{tab:modbus_features}
\centering
\scriptsize
\begin{tabular}{l l}
\toprule
\textbf{Feature} & \textbf{Type}  \\
Address & Network information \\
Function Code & Command Payload \\
Length of Packet & Network Information \\
Setpoint & Command Payload \\
Gain & Command Payload \\
Reset Rate & Command Payload \\
Deadband & Command Payload \\
Cycle Time & Command Payload \\
Rate & Command Payload \\
System Mode & Command Payload \\
Control Scheme & Command Payload \\
Pump & Command Payload \\
Solenoid & Command Payload \\
Pressure Measurement & Response Payload \\
CRC Rate & Network Information \\
Command Response & Network Information \\
Time & Network Information \\
Binary Attack & Label \\
Categorised Attack & Label \\
Specific Attack & Label \\
\bottomrule
\end{tabular}
\end{table}
During this time,
a total of 60048 attacks has been performed,
belonging to one of 35 kinds of attack.
These attacks are clustered into seven categories as listed in Table~\ref{tab:modbus_attacks}.
The complete list can be found in the work of \textit{Morris et al.}~\cite{Morris.2015}.
\begin{table}[h!]
\renewcommand{\arraystretch}{1.3}
\caption{Categories of Attacks in \textit{DS1}}
\label{tab:modbus_attacks}
\centering
\scriptsize
\begin{tabular}{l r l}
\toprule
\textbf{Abbreviation} & \textbf{Label} & \textbf{Description}  \\
Normal & 0 & Normal Behaviour \\
NMRI & 1 & Naive Malicious Response Injection \\
CMRI & 2 & Complex Malicious Response Injection \\
MSCI & 3 & Malicious State Command Injection \\
MPCI & 4 & Malicious Parameter Command Injection \\
MFCI & 5 & Malicious Function Code Injection \\
DoS & 6 & Denial of Service \\
Recon & 7 & Reconnaissance \\
\bottomrule
\end{tabular}
\end{table}

\subsection{\textit{OPC UA}-based Data Set \textit{DS2}}
\begin{figure}[h!]
\centering
\includegraphics[width=0.48\textwidth]{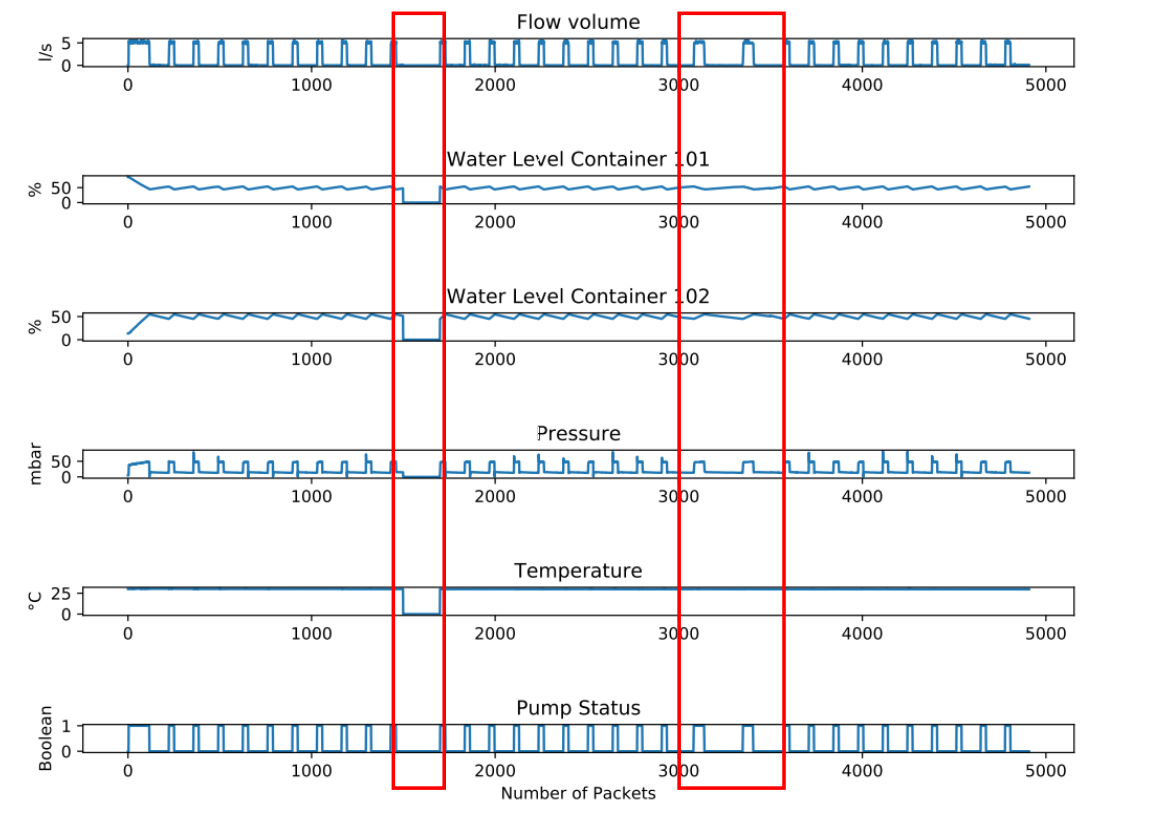}
\caption{Process Behaviour in \textit{DS2}}
\label{fig:opcua_normal_process}
\end{figure}

The \ac{opcua}-based data set analysed in this work is based on a real-world process environment introduced by \textit{Duque Anton et al.}~\cite{Duque_Anton.2019a}.
With the help of a \textit{Festo Didactic} environment,
the \textit{Festo Didactic MPS PA Compact Workstation},
a batch processing scenario is set up.
The water,
originally contained in \textit{Container 1},
is pumped into \textit{Container 2} by pump \textit{M102}.
Due to natural re-flow,
\textit{Container 2} is drained over time.
After the water level falls below a threshold value with a hysteresis,
the pump is restarted and pumps water from \textit{Container 1} into \textit{Container 2} again.
This normal behaviour is shown during the packets 0 to \numprint{1500},
from packet \numprint{1800} to packet \numprint{3000} and after packet \numprint{3500} in Figure~\ref{fig:opcua_normal_process}.
The attacks are from packet \numprint{1500} to packet \numprint{1800} and from packet \numprint{3000} to packet \numprint{3500},
marked by red frames.
In the first attack,
all sensor and actuator values are set to 0,
indicating an error in the process, 
causing a disruption due to necessary investigation and maintenance.
For the second attack,
the frequency of the process is cut in half.
This causes a change in the process,
still indicating activity.

\section{Algorithms Used}
\label{sec:algos}
In this work,  
\acp{svm} have been used to analyse the data sets as described in Section~\ref{sec:data} on packet basis.

\subsection{\acp{svm}}
\ac{svm} is a large margin classifier,
introduced by \textit{Boser et al.} in 1992~\cite{Boser.1992}.
Two classes of instances of dimensionality $n$ are divided by an $n+1$ dimensional hyperplane in a way that each instance has the maximal possible distance from the classifying hyperplane.
The instances are noted as tuples,
shown in (\ref{eq:svm_data})~\cite{Cortes.1995}.
\begin{equation}
\label{eq:svm_data}
\begin{split}
(x_{i}, y_{i}), i = 1, ..., m, y \in \{-1, 1\}
\end{split}
\end{equation}
$x$ is a vector describing an instance of data in an $n$-dimensional feature space.
$y$ describes the attribution of the instance as belonging to one of two classes,
while $m$ is the number of instances.
First,
the \ac{svm} is trained with a labeled set of instances.
It is a supervised classification method,
meaning the training set needs to contain information about the correct classes.
After training,
the attribution of the test and productive data is performed with the signum-function as shown in (\ref{eq:svm_attr}).
$w$ is the normal vector of the separator hyperplane,
$b$ is the offset from the hyperplane.
\begin{equation}
\label{eq:svm_attr}
\begin{split}
y_{i} = sgn(w, x_{i} - b)
\end{split}
\end{equation}
When applying \acp{svm},
obtaining a linear hyperplane to divide the data set is desirable.
However,
some data sets might not allow that.
In these cases,
the so-called \textit{kernel trick} can be applied~\cite{Cortes.1995}.
This \textit{kernel trick} can be used to map the input data space onto a higher dimensional feature space in a non linear fashion.
The higher dimensional data space then can be divided by a hyperplane in a linear fashion.

\subsection{\textit{Random Forests}}
\textit{Random Forests} are collections of \textit{Decision Trees},
binary classifiers consisting of one \textit{root} node,
several internal \textit{split} nodes and \textit{leaf} nodes that are used to classify events~\cite{Breiman.2001}.
The classification of the \textit{Random Forest} is done per majority vote of all \textit{Decision Trees}.
\textit{Random Forests} are robust to over-fitting and converge quickly,
making them applicable in a variety of use cases.
In addition to classifying data,
\textit{Random Forests} provide means to determine the relevance of features on the classification result.
This is done by measuring the decrease in accuracy if a feature is not considered for classification and the \textit{Gini} index respectively.
The \textit{Gini} index is a measure for the pureness of a data set that is split according to a certain feature~\cite{Rokach.2005}.
A high decrease in \textit{Gini} index indicates a high importance of the feature for a correct classification.
The decrease in accuracy is used to determine the feature relevance in Section~\ref{sec:eval}.

\section{Evaluation}
\label{sec:eval}
In this section,
the algorithms mentioned in Section~\ref{sec:algos} are applied to the data sets introduced in Section~\ref{sec:data}.
First,
the process of feature extraction is discussed and the features employed in anomaly detection are presented.
Furthermore,
the handling of missing data is addressed.
After that,
the performance of the respective algorithms is presented.

\subsection{Feature Selection}
Feature selection is useful in cases of large datasets where some features can be omitted without compromising on the accuracy of the classification model,
thereby resulting in reduced training time.
This is done by finding relationships in the data and analysing various processes that help in extracting such relations.
Selecting features that are capable of distinguishing between malicious and non-malicious instances is an important prerequisite to anomaly detection.
There are hardly formal methods to identify insightful features.
However,
there are approaches to identify the impact of a feature on the outcome,
e.g. with the importance score of \textit{Random Forests}~\cite{zhang2005network}.
In this work,
the features of \textit{DS1} have been analysed a priori.
These features and their relevance according to the importance score of \textit{Random Forests} are shown in Table~\ref{tab:ds1_feature_relevance}.
\begin{table}[h!]
\renewcommand{\arraystretch}{1.3}
\caption{Relevance of Individual Features in \textit{DS1}}
\label{tab:ds1_feature_relevance}
\centering
\scriptsize
\begin{tabular}{l l}
\toprule
\textbf{Feature} & \textbf{Relevance}  \\
Pressure & 0.177116 \\
Length & 0.134636 \\
CRC & 0.089498 \\
Cycle & 0.082905 \\
Reset & 0.082821 \\
Setpoint & 0.081545 \\
Function & 0.080561 \\
Gain & 0.068976 \\
Deadband & 0.060087\\
Rate & 0.039752\\
Command & 0.036751\\
System & 0.022929 \\
Pump & 0.021371 \\
Control & 0.010001 \\
Solenoid & 0.009587 \\
Address & 0.001463 \\
\bottomrule
\end{tabular}
\end{table}
Two variations of random forest classifiers were employed:
An ensemble with \ac{svm} by limiting the number of features for training the selection model and a classification of the full set of features for each dataset on plain random forest classifier.
On one hand,
a reduction of training time in using the ensemble classifier was observed.
On the other hand,
the random forest-only classifier resulted in a higher accuracy rate.
A perfect classification with no false positives or negatives is obtained for \textit{DS2}. 
By using random forests before training the \ac{svm} model for \textit{DS1}
a reduction of training time by \numprint{5810} seconds is observed while still obtaining an accuracy of \numprint{92.5}\%.
It is shown that the pressure as well as the length of a packet have a high impact on the classification,
as listed in Table~\ref{tab:ds2_feature_relevance}.
\begin{table}[h!]
\renewcommand{\arraystretch}{1.3}
\caption{Relevance of Individual Features in \textit{DS2}}
\label{tab:ds2_feature_relevance}
\centering
\scriptsize
\begin{tabular}{l l}
\toprule
\textbf{Feature} & \textbf{Relevance}  \\
Water Temperature & 0.241810 \\
Water flow volume & 0.201617 \\
Water level container 1 & 0.160073 \\
Water level Container 2 & 0.156903 \\
Water pressure & 0.155168 \\
S111 & 0.034989 \\
S113 & 0.034306 \\
Pump running & 0.008172 \\
Pump status & 0.004048 \\
S112 & 0.002526 \\
B114 & 0.000387 \\
Ball valve acknowledge & 0.000000 \\
\bottomrule
\end{tabular}
\end{table}
The next five features are similarly important,
followed by two slightly less important features,
gain and deadband.
The remaining features are less important.
The data sets were split into training and testing data in a relation of 80\% to 20\%.
This relation is consistent for normal and malicious events in order to effectively train the classifiers on both classes.
Furthermore,
the data was pre-processed:
A \ac{pca} was applied in using \textit{Random Forests} and a zero mean scaling was used for \acp{svm}.

\subsection{Handling Missing Data}
Missing values in \textit{DS1} constitute 40\% of the entire dataset.
The sizeable amount makes it crucial to handle the data efficiently so as to not deteriorate the detection rate of the classifier.
Missing data is randomly spread across the features and is not specific to any one feature or the category of attack a packet belongs to,
thus classified as \ac{mcar}~\cite{schafer1998multiple}.
The missing data comes from the characteristic of \textit{Modbus} packets to not transfer all data values automatically,
but only those requested.
In order to obtain data at regular time points,
this lack of information on several time points is impacting the feature selection.
As it is safe to assume that the data does not behave erratically in between polling for the given data, 
interpolation with time method is used for each feature of the dataset to estimate missing values as this method is efficient and most widely used~\cite{kreindler2016effects}.
The results indicate that interpolating missing values does not downgrade the accuracies of the different training models tested.
Since an attack may or may not impact all features at a single point in time,
each feature is considered to be an individual event.
For instance,
considering an MSCI attack,
an attack on the pump triggers random changes to the state of the pump,
whereas a solenoid attack randomly changes the state of the solenoid~\cite{Morris.2015}
The impact of an attack may eventually trickle down to other features in time.
However,
in this work,
only the point in time during which an attack occurs is discussed.
In \textit{DS2},
the dimensionality is lower.
From manual inspection of the features,
a few correlations are inferred.  
To get an idea of the trends in all attacked records,
manual analysis is performed on the dataset.
The following direct correlations between a feature and an attack are drawn:
\begin{itemize}
\item Attacks have an influence of the following attributes: Water flow, water level in both containers, water pressure and water temperature
\item Pump status remains constant during an attack: 0 during the first attack and 1 during the second attack
\item \textit{S111} is 0 during the first attack and 1 during the second attack and normal operation
\item \textit{S112} 0 during the first attack and during normal operation
\item \textit{S113} is 0 during the first attack
\end{itemize}
These observations are supported by the feature importance scores derived from random forest classifier.
The features as shown in Table~\ref{tab:ds2_feature_relevance}.
Water temperature is the most relevant feature with about 25\%.
The following four features are important as well,
while the remaining features are significantly less relevant.

\subsection{Performance of the Algorithms}
Since both the datasets are highly imbalanced with only 21.9\% of events being attacks in  \textit{DS1} and 14.3\% in \textit{DS2},
modifications in support vector machine are employed to appropriately tackle the class imbalance problem.
An 80/20 split for training and testing data was chosen for the data sets.
For \textit{Random Forests},
\ac{pca} was applied as a pre-processing means while a zero mean scaling was performed for \acp{svm}.
For \textit{DS1},
cost sensitive learning for \ac{svm} modeling using \ac{svm}-weight is implemented~\cite{tang2009svms}.
When the classification model is assigned \ac{svm}-weight,
it assigns a larger penalty value to \acp{fn} than to \acp{fp}.
The training time for \ac{svm}-weight is the same as the training time required for \ac{svm} and is of order $\mathcal{O}((Np+Nn)^3)$,
where $Np$ are positive samples and $Nn$ are negative samples.
However overweighting the classifier increases the training time.
There are other data processing techniques such as under-sampling where only a subset of majority class is used for training along with minority class~\cite{drummond2003c4},
and oversampling where samples of minority class are replicated.
Since oversampling increase the dataset size and under-sampling could result in decrease of meaningful time dependent trends,
the cost metric of the classifier using recursive validation with \textit{Gridsearch}~\cite{GridSearch.},
a Python library,
was performed in the course of this work.
An overview of the performance of \acp{svm} on the data sets is provided in Table~\ref{tab:svm_performance}.
\begin{table}[h!]
\renewcommand{\arraystretch}{1.3}
\caption{Performance of \acp{svm}}
\label{tab:svm_performance}
\centering
\scriptsize
\begin{tabular}{l r l}
\toprule
\textbf{Metrics} & \textbf{\textit{DS1}} & \textbf{\textit{DS2}}  \\
Accuracy & 92.5\% & 90.8\% \\
Precision & 78.2\% & 90.4\% \\
Recall & 93.6\% & 99.9\% \\
F1 Score &  85.2\% & 94.9\%  \\
\bottomrule
\end{tabular}
\end{table}
With highly skewed distribution of the data,
evaluation of the classifier based only on accuracy is not sufficient.
Additional metrics,
such as precision,
recall and F1-score,
are employed to evaluate the classification ability.
The confusion matrix of each trained classifier is generated to compare predicted output labels to the existing ground truth.
The confusion matrix consists of the following metrics:
\ac{tp},
i.e. the number of correct classification of attacks,
\ac{tn}
i.e. the number of correct classification of normal instances,
\ac{fp},
i.e. the number of incorrect classification of normal instances as attacks,
and \ac{fn},
i.e. the number of incorrect classifications of attacks as normal instances.
In addition to \acp{svm},
\textit{Random Forests} were employed in order to classify the relevance of features and to detect attacks in the data set.
A summary of the evaluation is shown in Table~\ref{tab:eval_ov}.
\begin{table}[h!]
\renewcommand{\arraystretch}{1.3}
\caption{Summary of the Evaluation of Individual Methods}
\label{tab:eval_ov}
\centering
\scriptsize
\begin{tabular}{l r r c r r}
\toprule
& \multicolumn{2}{c}{\textbf{\textit{DS1}}} & \phantom{a} & \multicolumn{2}{c}{\textbf{\textit{DS2}}} \\
\cmidrule{2-3} \cmidrule{5-6}
\textbf{Method} & \textbf{Acc(\%) } & \textbf{Exec. time (s)} & & \textbf{Acc(\%)}& \textbf{Exec. time (s)}\\
\ac{svm} & 92.53 & 11712 & & 90.81 & 0.019\\
RF & 99.84 & 281 & &  99.98 & 52.31  \\
\bottomrule
\end{tabular}
\end{table}
The columns on the left describe \textit{DS1} while the columns on the right describe \textit{DS2}.
For both the datasets,
full feature models of \textit{Random Forest} classifiers worked well with accuracies of 99.7\% and 99.9\% for \textit{DS1} and \textit{DS2} respectively.
On setting a threshold to filter out least relevant features,
the accuracy of both the datasets reduced:
To 91\% when model was trained with  9 most important features for \textit{DS1} and 92\% when model was trained with ten most important features from the derived rankings. 
The first data set,
\textit{DS1} consists of \textit{Modbus}-based traffic of a simulated process of a gas pump.
\textit{Random Forests} reach a significantly higher accuracy than \acp{svm} while exhibiting a more linear time behaviour.
As they converge quickly,
the take less time for the larger data set \textit{DS1}.

\section{Conclusion}
\label{sec:concl}
In this work,
two data sets captured in industrial environments are analysed for attack-based anomalies.
\acp{svm} are used to detect attacks of seven different categories and 35 different subtypes.
With accuracies and F1-scores of up to 92.5\% and 85.2\% respectively,
this approach is promising.
Lots of missing data in this data set do not impact the performance of the algorithms.
The second data set, 
\textit{DS2} consists of \ac{opcua}-based traffic derived from a real-world hardware,
a \textit{Festo Didactic MPS PA Compact Workstation}.
In this case, 
one class \acp{svm} are used,
as only two instances of attacks are present.
This leads to slightly worse accuracy of 90.8\%.
However,
due to an almost perfect recall,
the f1-score is performing pretty well with 94.9\%.
In this work,
handling of missing data as well as feature selection are addressed,
as real-world data often lacks features for some instances and contains noise and irrelevant information.
\textit{Random Forests} provide means to calculate the significance of individual features.
They can also be employed to detect anomalies in the data set,
with satisfactory results.
Additionally,
approaches such as \ac{pca} decrease the feature space and provide the most relevant features.
As attacks on industrial environments are occurring more frequent and becoming more critical in their effects,
effective detection is crucial.
Methods of machine learning,
such as \ac{svm} and \textit{Random Forests},
are capable of enhancing the detection capabilities of common industrial \acp{ids}.
Machine learning-based approaches profit from a limited state set in industry as well as a large amount of training data in a given environment,
as industrial environments produce loads of data.
The algorithms presented in this work do detect between 90\% and 95\% of the attacks in the data. 
This will,
in practical application,
leave undetected attacks in industrial networks and thus pose security threats.
Additional means are required to enhance the level of security~\cite{Duque_Anton.2019b, Duque_Anton.2019d, Duque_Anton.2019e}.

\bibliographystyle{IEEEtran}
\bibliography{literature}

\end{document}